\documentclass[10pt,prb,aps,twocolumn,showpacs,floatfix,superscriptaddress]{revtex4}

\usepackage{graphicx}
\usepackage{amssymb}
\usepackage{subfigure}
\usepackage{color}
\usepackage{amsmath} 
\begin{document}

\title{Two-orbital quantum spin model of magnetism in the iron pnictides}

\author{Chen Liu}
\affiliation{Department of Physics, Boston University, 590 Commonwealth Avenue, Boston, Massachusetts 02215, USA}

\author{Dao-Xin Yao}
\affiliation{State Key Laboratory of Optoelectronic Materials and Technology, Sun Yat-Sen University, Guangzhou, China}

\author{Anders W. Sandvik}
\affiliation{Department of Physics, Boston University, 590 Commonwealth Avenue, Boston, Massachusetts 02215, USA}

\begin{abstract}
We study a two-orbital spin model to describe $(\pi,0)$ stripe antiferromagnetism in the iron pnictides. The ``double-spin'' model 
has an on-site Hunds's coupling and inter-site interactions extending to second neighbors (inter- and intra-orbital) on the square lattice. 
Using a variational method based on a cluster decomposition, we optimize wave functions with up to $8$ cluster sites (up to $2^{16}$ variational 
parameters). We focus on the anomalously small ordered moments in the stripe state of the pnictides. To account for it, and large variations among 
different compounds, we show that the second-neighbor cross-orbital exchange constant should be ferromagnetic, which leads to ``partially 
hidden'' stripe order, with a moment that can be varied over a large range by small changes in the coupling constants. In a different parameter 
region, we confirm the existence of a canted state previously found in spin-wave theory. We also identify several other phases of the model.
\end{abstract}

\date{October 4, 2011}
\pacs{74.70.Xa, 75.10.Jm, 75.50.Ee, 75.30.Et}

\maketitle

\section{Introduction}

Like the cuprate high-$T_{\rm c}$ superconductors, the more recently discovered iron pnictide superconductors \cite{iron1,iron2} 
also exhibit interesting magnetic properties.\cite{mag1,mag2,sdw,reduced,lynn09} While the antiferromagnetism of the Mott-insulating 
parent compounds of the cuprates can be very well accounted for by the two-dimensional (2D) $S=1/2$ Heisenberg model,\cite{Mreview} there 
is no such simple reference point for the Fe pnictides, in which the electrons are always itinerant. Quantum spin models have nevertheless been 
employed  \cite{sp1,sp2,si,canted,dis,rod10,schmidt}  to describe some of their intriguing magnetic properties, as more manageable alternatives to 
t-J \cite{tj} or Hubbard \cite{daghofer08,schickling} models of itinerant electrons. In contrast to the $(\pi,\pi)$ (N\'eel) order of the 
cuprates, most of the pnictides exhibit $(\pi,0)$ ``stripe'' order, which can be achieved with frustrated spin models.\cite{j1j2} An on-going 
theoretical challenge is to explain,\cite{si,yin11} in a general sense, the ordered moments, which are often anomalously small but vary 
considerably among different compounds.\cite{mag1,mag2,sdw,reduced,lynn09}

We here address this issue, as well as other generic aspects of multi-orbital magnetism, using a model with two electronic spins per lattice 
site. While the neglect of the charge degrees of freedom will clearly also affect the magnetic properties to some extent, it is still interesting 
and important to ask the question of whether a correct effective low-energy description of the magnetic properties of the Fe pnictides is possible within 
such a simplified spin-only model. The two-orbital spin system should be much more realistic in this regard than the often used models with single $S=1/2$ 
or $S=1$ spins on the sites.\cite{sp1,sp2,schmidt} We will include in the two-spin description an on-site Hund's coupling, as well as inter- and intra-orbital 
exchange between nearest and next-nearest-neighbor spins. 

Studying this highly frustrated quantum spin system with a cluster-mean-field method (a variational cluster-product-state ansatz where the wave
function on small clusters is optimized fully), we find a very rich phase diagram, including a regime in which the behavior seems appropriate
for describing the small-moment stripe antiferromagnetism of the pnictides.

\section{Double-spin model}

The valence electrons in the Fe pnictides reside predominantly in Fe $d$ orbitals. In a simplified picture suitable for some 
compounds, the low-energy bands \cite{marzin08} can be reproduced using only $d_{xz}$ and $d_{yz}$ orbitals \cite{raghu08} (although in some cases 
other orbitals may also be needed \cite{simple1,si}). In undoped systems there are two electrons per Fe in this case. Despite the electron itineracy, 
it has been argued that the magnetic properties are still correctly maintained upon further reducing the two-orbital Hubbard model to a 
``double-spin'' model (DSM),\cite{si,seo08} where two localized $S=1/2$ spin degrees of freedom per lattice site are retained. 
The ``A'' and ``B'' spins are coupled to each other intra-site by a ferromagnetic Hund's coupling $J_H$, and significant inter-site 
couplings should extend up to second-nearest neighbors on the square lattice of iron ions. These interactions originate 
from superexchange through As ions.\cite{yildrim08} 

The spin-only description of course neglects some important aspects of the pnictides.
One appealing proposal to explain their small and highly varying magnetic moments is a ``semi-localized'' picture, in which the electrons
on one orbital are essentially localized, while those on the second orbital form an itinerant band.\cite{wu} The two-orbital spin-only 
description would then amount to a effective spin-localized description also of the itinerant electrons. Whether or not such a model can still correctly 
capture the low-energy magnetic properties is an interesting and important generic question. The aim of our work reported here is to establish
some aspects of the phase diagram of the DSM more completely than in past studies,\cite{si,canted,dis,rod10} as a starting point for investigating 
the low-energy physics in various phases and the nature of the quantum phase transitions between the different types of ground states.

\begin{figure}
\centering
\includegraphics[width=6.75cm, clip]{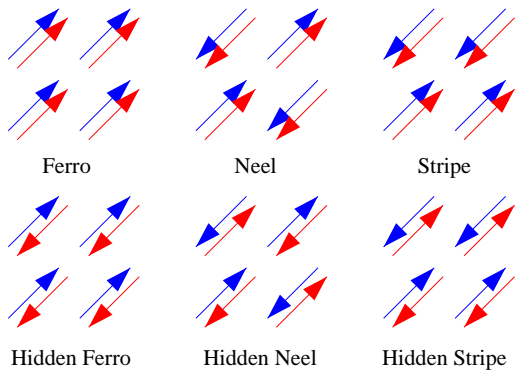}
\vskip-1mm
\caption{(Colored online) Six phases with collinear order appearing in the DSM. In the standard ferrormagnetic, N\'eel and stripe
phases (upper row), the two spins on the same site are fully aligned. In the corresponding hidden-order phases (bottom row) there 
is no net on-site moment, due to the two spins on the same site being anti-parallel.}
\vskip-1mm
\label{fig1.5}
\end{figure}

Using a notation with a site index $i \in \{1,\ldots,N\}$ and an orbital index $\alpha \in \{A,B\}$ on the  $S=1/2$ spin operators $\mathbf{S}_{i\alpha}$, 
the DSM in the general form considered here is defined by the Hamiltonian
\begin{eqnarray}
H&=&\sum_{\langle i,j \rangle}\sum_{\alpha,\beta} J_1^{\alpha\beta}\mathbf{S}_{i\alpha}\cdot\mathbf{S}_{j\beta} \label{ham1} \\ 
&~~~+&\sum_{\langle \langle i,j \rangle \rangle}\sum_{\alpha,\beta} J_2^{\alpha\beta}\mathbf{S}_{i\alpha}\cdot\mathbf{S}_{j\beta}-
J_H\sum_i\mathbf{S}_{i,A}\cdot\mathbf{S}_{i,B},\nonumber
\end{eqnarray}
where $\langle i,j \rangle$ and $\langle \langle i,j \rangle \rangle$ denote nearest- and next-nearest-neighbor pairs of sites, respectively, on the 
2D square lattice. Here we count the A-B cross-terms only once, i.e., $\alpha \ge \beta$ in the sum over orbitals. This model has up until now been 
studied only in a very limited range of its parameter space.\cite{canted,dis,rod10} We here report much more extensive investigations, using a 
variational cluster-product wave-function ansatz. 

Our primary goal is to determine whether the DSM can provide a semi-quantitative, universal description of the magnetism of the pnictides and insights into their 
effective magnetic interactions. To address the issue of typically small but highly varying ordered moments in the $(\pi,0)$ stripe phase, we take the view 
that the wide range of moments should originate from a sensitivity of the system to details of the effective spin couplings. It should therefore be 
possible to achieve small moments for a wide range of parameters in the DSM. The sensitivity should not require the proximity of a quantum-critical point,
which seems to require too much fine-tuning to be generic. Note that the variations in the moments are not only large among different members of the Fe 
pnictide family, but the moment of a given system can also be sensitive to pressure.\cite{ma} 

We also emphasize that the DSM is not just of interest in the context of the Fe pnictides, but also represents an intriguing and important quantum spin 
model in its own right, with potential applications to other multi-orbital quantum magnets. It is interesting, e.g., to observe how the DSM evolves into 
an $S=1$ system with increasing Hund's coupling. More generally, we are interested in the quantum many-body states that can form in this
system due to the two-orbital spin physics, beyond the known ground states of single-orbital $S=1/2$ systems and $S=1$ ($J_H \to \infty$) 
systems. 

Using the variational cluster mean-field method with fully optimized cluster wave functions, as described in more detail below, in Sec.~\ref{method},
we find several ordered magnetic phases in the DSM, including ones previously identified and also some that had not been noted in the earlier studies 
of the double-spin model in a more limited parameter space.\cite{canted,dis,rod10} To illustrate some of the complexity of the phase diagram, Fig.~\ref{fig2} 
shows simple collinear ordered and ``hidden-order'' phases that appear in the model in certain limits (that we will discuss in detail further below). 
Other, more interesting non-collinear phases with highly non-trivial quantum fluctuations appear when tuning parameters between these limits.

\begin{figure}
\includegraphics[width=4.75cm, clip]{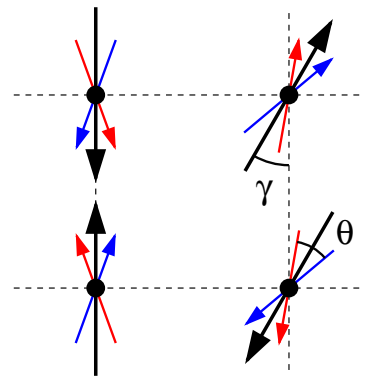}
\vskip-1mm
\caption{(Color online) Illustration of average $A$ (red), $B$ (blue), and total spins (black) in a unit cell of the canted state. 
The canting angle $\gamma$ is obtained by rotating every second column (here the right one) with respect to the perfect N\'eel state 
$(\gamma=0)$, so that $\gamma=\pi$ in the stripe state. $\theta$ is the angle separating the A and B directions.}
\label{fig1}
\vskip-1mm
\end{figure}

Being highly frustrated when at least some of the interactions are antiferromagnetic, the DSM is much more complicated and difficult to study 
reliably than the standard Heisenberg model for the cuprates. Having two coupled spins per site, it is also much more complex than the $S=1/2$ 
$J_1$-$J_2$ Heisenberg model.\cite{j1j2} Spin-wave theory \cite{canted,rod10} and exact diagonalization \cite{dis} have been 
used to study some regions of its parameter space. Retaining only $J_H$ and the orbit-diagonal couplings $J_1^{AA}$ and $J_2^{BB}$, Manousakis 
discovered a canted phase with a $2\times 2$ unit cell (illustrated in Fig.~\ref{fig1}),\cite{canted} in addition to the N\'eel and stripe 
states. He also suggested a non-magnetic phase for large $J_H$. Rodriguez and Rezayi \cite{dis,rod10} focused on completely different parameters, 
with only cross-orbital interactions $J_1^{AB}$ and $J_2^{AB}$ and weak Hund's coupling. States with ``hidden order'' then appear 
(Fig.~\ref{fig2}) where N\'eel or ferromagnetic order is present but macroscopically invisible because spins on the same site are anti-parallel 
(overcoming $J_H$) and lead to no net moment (if the effective $g$-factors of the two electrons are equal). These parameters do not seem appropriate 
for the pnictides, however, and the kinds of hidden order also seem inappropriate, because it does not correspond to the commonly observed
$(\pi,0)$ stripes in the Fe pnictides.

The full phase diagram of the DSM (\ref{ham1}) has not been studied so far. Even in the regions already studied, the methods used may not be 
completely reliable, due to the inherent difficulties with spin-wave theory when the quantum fluctuations are significant and the small lattices in 
exact diagonalization. It is therefore important to investigate the DSM for a wider range of parameters and using alternative methods.

\section{Cluster-variational method}
\label{method}

The variational approach used here is based on factoring of the ground state of the system on the infinite lattice into a product of 
cluster states, 
\begin{equation}
|\Psi\rangle = \prod_c |\psi_c\rangle, 
\end{equation}
where $\psi_c$ is the wave function of cluster $c$. We here use the standard basis of z-component spin eigenstates and write the state of 
a cluster of $n$ sites as
\begin{equation}
|\psi_c\rangle = \sum_\sigma c_\sigma |\sigma\rangle,~~~|\sigma\rangle = |S^z_1,\ldots S^z_n\rangle.
\label{psicdef}
\end{equation}
Imposing the constraint of all the cluster wave functions being identical corresponds to periodicity with a maximal unit cell given by the cluster used and 
mean-field cluster boundaries,\cite{mata88} i.e., the interactions between different clusters factor into expectation value taken over two independent
clusters. Since all clusters have the same wave function, these expectation values just depend on a single cluster. In principle one may also extend the scheme 
to a product of different, collectively optimized cluster states (to allow for, e.g., magnetic order with larger periodicity), but we will not consider 
this generalization here.

\subsection{Energy minimization}

Using clusters of size $N=2\times 2$ and $2\times 4$, we have fully optimized the wave function (which for the DSM has $4^n$ coefficients) 
by minimizing the total energy,
\begin{equation}
E=\frac{\langle\psi |H|\psi\rangle}{\langle \psi | \psi \rangle}=\frac{\sum_{\sigma\tau} c_\sigma c_\tau \langle \tau|H|\sigma \rangle }{\sum_\sigma {c_\sigma}^2},
\label{energy}
\end{equation}
with respect to the parameters $c_\sigma$. To minimize $E$, we also compute its derivatives with respect to all the parameters. It is useful
to divide the energy into its diagonal (dia) and off-diagonal (off) parts, e.g., for the standard Heisenberg model (single-orbital) these are
given by
\begin{eqnarray}
E_{\rm dia} & = & \sum_{\langle i,j\rangle} \langle S^z_iS^z_j \rangle,\\
E_{\rm off} & = & \hbox{$\frac{1}{2}$}\sum_{\langle i,j\rangle} \langle S^+_iS^-_j+S^-_iS^+_j \rangle,
\end{eqnarray}
and the generalization to the DSM energy is trivial. The general expressions for the diagonal and off-diagonal contributions to the energy derivatives are
\begin{eqnarray}
\frac{\partial E_{\rm dia}}{\partial c_\sigma}&=&\frac{2c_\sigma(E_{\sigma\sigma}-E_{\rm dia})}{\sum_\tau c_\tau^2}, \\ 
\frac{\partial E_{\rm off}}{\partial c_\sigma}&=&\frac{\sum_{\tau}c_{\tau}E_{\sigma\tau}-2c_\sigma E_{\rm off}}{\sum_\tau c_\tau^2},
\label{derivative}
\end{eqnarray}
where $E_{\sigma\tau}=\langle \tau | H | \sigma \rangle$, with $H$ is restricted to a single cluster and the mean-field decoupled inter-cluster
interactions discussed above are included.

The cluster-variational (CV) method outlined above is in practice very similar to the ``hierarchical mean-field theory'' \cite{hmf1,hmf2,hmf3} (producing identical 
results for systems such as the $J_1$-$J_2$ model \cite{mf1}), which in the past has been applied to several frustrated spin models. Our method of constructing
the wave function by direct energy minimization is different, however. We do not make any assumptions regarding the form of the cluster wave function, treating
all the coefficients in (\ref{psicdef}) as completely independent variational parameters. We normally do not implement any symmetries (which would reduce the 
number of variational parameters) in order to not potentially miss any states. In principle symmetries can also be easily incorporated in our scheme. Note that 
the magnetization 
\begin{equation}
m_z=\sum_{i=1}^n (S^z_{i,A}+S^z_{i,B})
\end{equation}
on the cluster is not conserved in general, but when studying the colinear states, such as the N\'eel and stripe states, one can restric the wave function
to the sector $m_z=0$ to reduce the number of parameters. In unrestricted simulations the optimization does not necessarily find this subspace, because the
spin-rotational symmetry is not broken explicitly (only spontaneously) and a state with $m_\gamma=0$ in any direction $\gamma$ can be generated. All these
solutions are degenerate and this is manifested in our calculations.

We find the optimum using a combination of the steepest-decent method and a stochastic approach where only the signs of the derivatives are used,\cite{lou07} 
with the latter useful at the initial stage where it is important to avoid getting stuck in local minimums, and the latter working well at the final stage where 
the solution is close to optimal. For the models considered here, we have not encountered any significant problems with this optimization scheme, even when the 
number of variational parameters is as large as $2^{16}=65536$ (for a $4 \times 4$ single-orbital $S=1/2$ Heisenberg model or a $4\times 2$ cluster of the DSM).

\subsection{Order parameters}

We here focus on the magnetic structure and define the N\'eel ($M_1$) and stripe ($M_2$) order parameters
\begin{eqnarray}
\mathbf{M}_1 & = & \frac{1}{N}\sum_{i=1}^N \mathbf{S}_i(-1)^{x_i+y_i}, \label{m1def}\\
\mathbf{M}_2 & = & \frac{1}{N}\sum_{i=1}^N \mathbf{S}_i(-1)^{x_i}, \label{m2def}
\end{eqnarray}
where $(x_i,y_i)$ are the integer-valued site coordinates. Naturally, these quantities as well can be expressed using only a single
cluster in the CV scheme.

The CV ansatz can break spin and lattice symmetries, but we do not impose how they are broken. We therefore compute the magnitudes of the full vector quantities
(\ref{m1def}) and (\ref{m2def}). In the case of stripes, $M_2$ corresponds to$(\pi,0)$ order (vertical stripes), but we also compute the $(0,\pi)$ order parameter 
and use whichever one (if any) is nonzero. In addition to the order parameters we also compute the canting angle $\gamma$ and the on-site separation angle $\theta$ 
defined in Fig.~\ref{fig1}. In calculations with $2\times 4$ clusters we compute the magnetic properties on the central $2\times 2$ plaquette.

\section{Results}

The magnetic structure within linear spin-wave theory is obtained by minimizing the classical energy, with quantum fluctuations only reducing (or possibly 
destroying) that order. We will compare some of our results with spin wave theory. The CV method includes local quantum fluctuations within the clusters, and 
by using different cluster sizes we can check the stability of any states found. In practice we are of course limited to very small clusters, and cannot in 
general carry out completely unbiased extrapolations. The method nevertheless goes well beyond classical or standard mean-field approaches. The CV method 
should also lead to faster convergence with the cluster size in ordered phases (if this order is commensurate with the cluster structure) than standard 
exact diagonalization techniques. We here discuss several aspects of the phase diagram, starting with the case of $J_H \to \infty$ (the spin-$1$ Heisenberg 
model) and then several other phases obtaining for finite $J_H$, including the parameter region we believe is appropriate for the pnictides.

\subsection{The $S=1$ limit}

\begin{figure}
\includegraphics[width=7cm, clip]{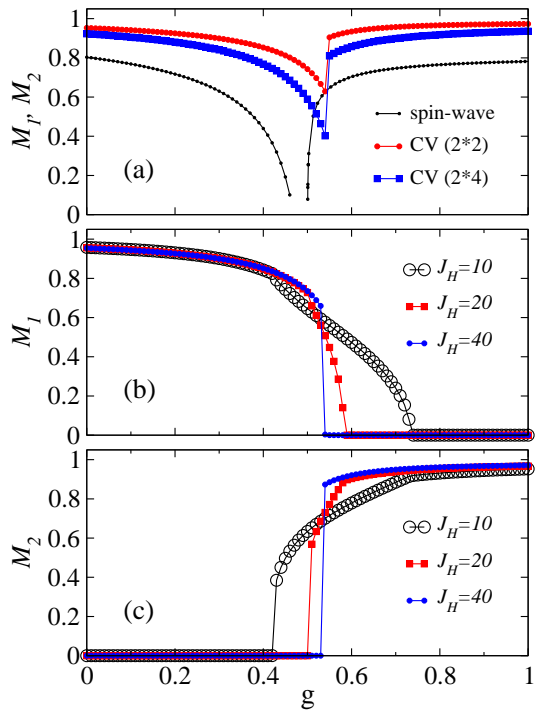}
\vskip-1mm
\caption{(Color online) (a) Neel ($M_1$) and stripe ($M_2$) order parameters of the $J_H=\infty$ DSM when the only non-zero parameter is $g=J_2^{BB}/J_1^{AA}$. 
The relevant order-parameter is $M_1$ (N\'eel) for $g\agt 0.5$ and $M_2$ (stripe) for $g \alt 0.5$. (b),(c) Dependence of the order parameters on $g$ for
three different Hund's couplings (using $2\times 2$ clusters). Both $M_1$ and $M_2$ are nonzero in the canted phase.}
\label{fig2}
\vskip-1mm
\end{figure}

We first discuss the limit $J_H \to \infty$, where the DSM reduces to a spin-$1$ system. In the sub-space of parameters considered 
by Manousakis \cite{canted} we define $g=J_2^{BB}/J_1^{AA}$ (and all other parameters are zero). Classically, there is then a first-order transition 
between the N\'eel and stripe states at $g=1/2$. As shown in Fig.~\ref{fig2}(a), within the CV approach with clusters of size $2\times 2$ and 
$2\times 4$, this transition moves only slightly, to $g\approx 0.55$. 

In linear spin-wave theory there is a narrow region at $g\approx 0.5$ where both ordered moments vanish (strictly speaking becoming negative, 
indicating a break-down of the approximation). Manousakis found that this behavior also persists for large but finite Hund's coupling, and, thus, 
that there is possibly a non-magnetic state intervening between the N\'eel and stripe phases. There is no explicit sign of such a state within the 
CV calculations, however. The N\'eel order drops significantly before the first-order stripe transition, and the drop increases with the cluster size. 
Such a drop is always expected in the presence of quantum fluctuations. In principle we cannot exclude that this behavior evolves with increasing
cluster size into two independent transitions with an intervening non-magnetic state in between. 

In the $S=1/2$ $J_1$-$J_2$ model it is well known that 
a nonmagnetic state exists for $0.45 \alt g \alt 0.6$.\cite{j1j2,capriotti00,jiang11}  This state is likely a valence-bond-solid,\cite{j1j2} although 
some calculations instead have indicated a spin-liquid.\cite{capriotti00,jiang11} A nonmagnetic state in the DSM, if it exists, should then be the $S=1$ 
version of such a state (which can exist \cite{read} for $S>1/2$ but is less studied than $S=1/2$). Considering that the $g$-window of the nonmagnetic 
phase is already small in the $S=1/2$ case, it should be very small indeed for $S=1$ (since it must vanish in the classical, $S\to \infty$ limit) and 
it is also possible that it is non-existent already for $S=1$. The transition should then be first order on generic symmetry grounds.

It is interesting to note that the location of the N\'eel to stripe transition is almost identical for the two cluster sizes, suggesting that 
the transition point is accurately converged. Moreover, the location also is close to the point obtained within a self-consistent harmonic 
approach.\cite{spin1} Note also that both ordered moments decrease with increasing cluster size. Spin-wave theory is known to be accurate for $g=0$ 
(where the model reduces to the standard $S=1$ Heisenberg antiferromagnet).\cite{Mreview} The CV result is about $12\%$ higher for these 
cluster sizes.

\subsection{Canted state}
\label{sec:canted}

\begin{figure}
\includegraphics[width=8cm, clip]{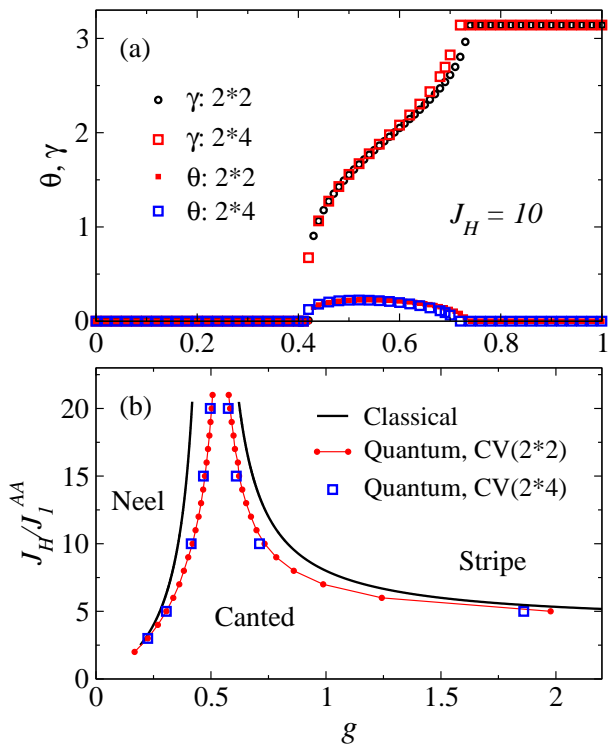}
\vskip-1mm
\caption{(Color online) Results for Manousakis' version of the DSM, where the only non-zero coupling is $g=J_2^{BB}/J_1^{AA}$ and $J_H$. 
(a) Evolution with $g$ of the canting and separation angles (defined in Fig.~\ref{fig1}) at $J_H=10$. (b) Boundaries in the $(g,J_H)$ plane
of the canted phase obtained in spin-wave theory (classical energy minimization) and within the CV method. In both (a) and (b) CV results are 
shown for both $2\times 2$ and $2\times 4$ clusters.}
\label{fig3}
\vskip-1mm
\end{figure}

The canted state found within spin-wave theory by Manousakis \cite{canted} also exists in our approach for intermediate $g$ when $J_H$ is finite. 
Examples of the evolution of the N\'eel and stripe order parameters with $g$ are shown in Fig.~\ref{fig2} for $2\times 2$ clusters. Fig.~\ref{fig3}(a) 
shows the angles $\gamma$ and $\theta$ for $J_H/J_1^{AA}=10$. The two different clusters give almost identical results. Note that the separation angle 
$\theta$ is nonzero only inside the canted state, where $0< \gamma < \pi$. The quantum phase transitions involving the canted state are clearly 
discontinuous for large $J_H$ but become more smooth as $J_H$ is reduced. 

The phase diagram in the $(g,J_H)$ plane is shown in Fig.~\ref{fig3}(b). The size of the canted phase is smaller than in spin-wave (classical) 
theory,\cite{canted} but the fact that the boundaries are barely shifted when increasing the cluster size from $2\times 2$ to $2\times 4$ suggests 
that the state indeed exists and the computed phase boundaries should be stable.

\begin{figure}
\includegraphics[width=8.4cm, clip]{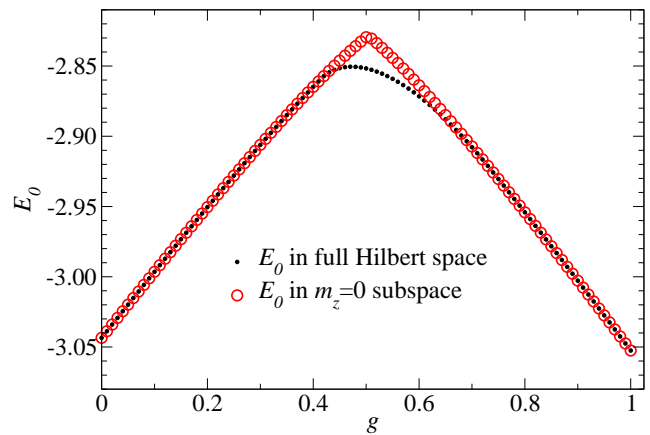}
\vskip-1mm
\caption{(Color online) The ground state energy of the DSM on the $2\times 2$ cluster versus the coupling ratio $g=J_2^{BB}/J_1^{AA}$ in Manousakis' 
version \cite{canted} of the DSM at $J_H=10$ in the $m_z=0$ subspace (red) and the whole Hilbert space (black). The energies deviate from each other in 
the non-collinear canted state [with the canting angle shown for the same parameters in \ref{fig3}(a)] but coincide exactly in the collinear states.}
\vskip-1mm
\label{fig3.5}
\end{figure}

In a collinear state, such as those in Fig.~\ref{fig2}, the ground state is in the sector of magnetization $m_z=0$. The canted state is a 
non-collinear (co-planar), however, and, thus, its ground state should mix different magnetization sectors (regardless of the quantization axis chosen). 
We can confirm this explicitly by comparing the CV optimized energy in the full Hilbert space of the cluster with that obtained in calculations restricted 
to $m_z=0$. An example of these two energies as a function of $g$ is shown in in Fig.~\ref{fig3.5}. The region in which the two energies are different 
(with the one from the unrestricted calculations naturally being lower when they differ) coincides with the window in which the canting angle is not $0$ or $\pi$.

\subsection{Modeling the Fe pnictides}

We now turn to our main objective of searching for parameters reproducing small and highly varying stripe-ordered
moments of the Fe pnictides. The stripe moment tends to be large in Manousakis' version of the DSM (as seen in Fig.~\ref{fig2}) and one cannot expect 
the neglected model parameters to be very small.\cite{yildrim08} We have investigated the full parameter space of the Hamiltonian (\ref{ham1}) extensively. 
In addition to the states discussed above and the hidden-order states studied in,\cite{dis} we have also identified a hidden stripe state 
(see Fig.~\ref{fig2}) and a transition between it and the normal $(\pi,0)$ stripes discussed above. 

\begin{figure}
\centering
\includegraphics[width=8.4cm, clip]{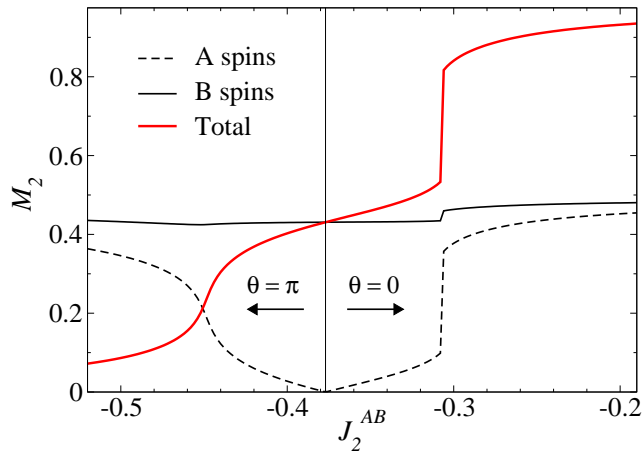}
\vskip-1mm
\caption{(Color online) Stripe moments in the DSM with couplings $J_H=2$, $J^{AA}_1=1$, $J^{AA}_2=0.5$, $J^{BB}_1=0.5$, $J^{BB}_2=0.5$, $J^{AB}_1=-0.7$, 
and variable $J^{AB}_2$, obtained with $2\times 2$ clusters. The separation angle $\theta$ changes from $\pi$ to $0$ at $J^{AB}_2 \approx -0.38$.}
\label{fig4}
\vskip-1mm
\end{figure}

Importantly, by considering couplings that are not symmetric with respect to the A and B spins, we can also achieve ``partially hidden'' stripe order, 
where the separation angle $\theta=\pi$ but the order parameter computed individually for the A and B sets of spins are unequal. It is then possible to tune 
the net stripe moment over a wide range of small to large values. This turns out to be the case only if the second cross-orbital coupling $J_2^{AB}$ is 
{\it ferromagnetic}, in which case this coupling favors A and B stripes out-of-phase with respect to each other (and also further stabilize the stripes). The Hund's 
coupling must be small to moderate; up to $2-4$ times the largest exchange constant. This is perhaps smaller than expected in the Fe pnictides,\cite{tj} 
but here it should be kept in mind that the bare $J_H$ (in an electronic model including the charge degrees of freedom) should be renormalized when 
reducing the itinerant description to a localized-spin model. With localized spins, the effects of the Hund's coupling is naturally stronger, and, therefore,
its bare value is renormalized down from its itinerant-electron value. All the other couplings can be antiferromagnetic, but a ferromagnetic $J_1^{AB}$ 
also helps to stabilize stripes. 

Fig.~\ref{fig4} shows an example of the evolution of the separation angle and the ordered moments as a function of $J_2^{AB}$ when 
all other couplings are held constant at physically reasonable \cite{yildrim08} values of order $1$ (using $J_1^{AA}$ as the energy unit). 
The canting angle jumps at $J^{AB}_2 \approx -0.38$, corresponding to a transition between partially hidden stripes and normal stripes. There is no 
discontinuity in $M_2$, however, because the individual moment of the A spins goes through zero at this point. The moment of the B subsystem 
is almost constant at $\approx 90\%$ of its maximal value $1/2$ (and can be reduced by adjusting other couplings). At $J^{AB}_2 \approx -0.31$ 
there is another interesting transition, where the stripe order jumps discontinuously with no qualitative change in the $(\pi,0)$ structure. 

\subsection{Other phases of the DSM}

\begin{figure}
\centering
\includegraphics[width=8.4cm, clip]{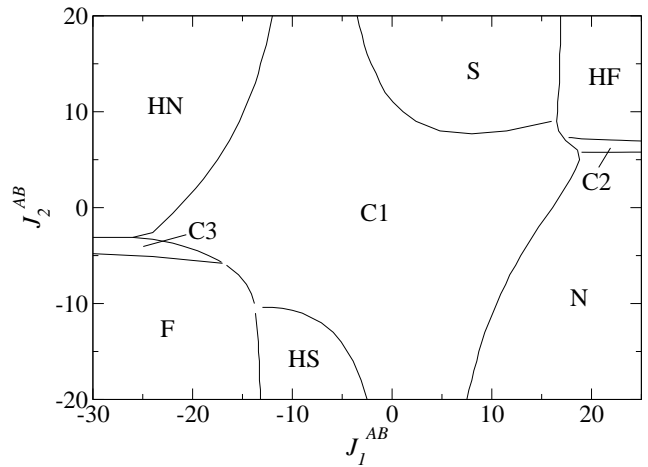}
\vskip-1mm
\caption{(Colored online) Phase diagram of the DS model in the subspace of varying $J_1^{AB}$ and $J_2^{AB}$ when all the other parameters are
held fixed at $J_H=5$, $J_1^{AA}=J_2^{BB}=10$ and $J_1^{BB}=J_2^{AA}=0$. The three collinear phases are ferromagnetic (F), stripe (S), and N\'eel (N). 
The hidden-order states are (where we have not drawn the boundaries that correspond to the coplanar partially hidden-ordered variants)
hidden (fully or partially) ferrormagnetic (HF), hidden (fully or partially) N\'eel, and hidden (fully or partially) stripes (HS). There
are also three canted states, C1, C2, C3, with spin structure discussed in the main text.}
\vskip-1mm
\label{fig4.5}
\end{figure}

The double-spin model studied here has six dimensionless parameters (coupling ratios) and a rich phase diagram in this space. To study phases beyond
those discussed above, we have carried out systematic scans throughout the large parameter space. Here we show illustrative results obtained when 
some of the parameters are fixed within the subspace previously considered by Manousakis (and here in Sec.~\ref{sec:canted}) in the canted phase; 
$J_H=5$, $J_1^{AA}=J_2^{BB}=10$ and $J_1^{BB}=J_2^{AA}=0$. We vary the cross-orbital couplings $J_1^{AB}$ and $J_2^{AB}$ to study how the canted state evolves. 
The resulting phase diagram is shown in Fig.~\ref{fig4.5}. In addition to the canted state existing around $J_1^{AB}=J_2^{AB}=0$ (marked as C1 in the figure), 
all six collinear phases illustrated in  Fig.~\ref{fig2} are also present here. Moreover, there are two other kinds of canted phases here as well, marked in the
figure as C2 and C3. These states interpolate between their adjacent collinear states, and correspond to the spin structure shown in Fig.~\ref{fig1} when 
switching the left and right spin of the bottom row (giving C2) or the top and bottom spin on the left column (giving C3).

\section{Summary and discussion}

We have presented a comprehensive study of the two-orbital spin model introduced by Si and Abrahams.\cite{si} Our results show that this system 
has a very rich phase diagram. In addition to the previously studied canted state interpolating between N\'eel and stripe antiferromagnetic 
states,\cite{canted} which we have shown to be stable against quantum fluctuations (although it is less extended than in spin-wave theory), there
are also other canted states interpolating between all the three ordered collinear states we have studied (ferromagnetic, N\'eel, and stripe). Moreover
the previously studied hidden-order phases also have partially-hidden generalizations, in which the contributions to the total ordered moment
from the two orbitals do not cancel completely. The partially hidden stripe state is of particular interest in the context of Fe pnictides,
reproducing the observed behavior of a sensitivity of the magnitude of the moment to small variations in the parameters, without necessarily
being close to a quantum-critical point. In this picture, the magnetism in the Fe pnictides is universally of the same type, but with different
degree of cancellation of the moments residing on the two orbitals.

One would expect antiferromagnetic couplings due to superexchange in the pnictides,\cite{si,seo08,yildrim08} but the effective ferromagnetic cross-orbital 
couplings that we have found here in association with the partially hidden stripe state may appear due to the itineracy of the electrons. The superexchange was 
calculated in Ref.~\onlinecite{yildrim08}, but only as effective $J_1$ and $J_2$ total-moment couplings, not separated into orbital-diagonal and cross contributions. 
A Hund's-like mechanism favoring parallel spins of A and B band electrons could come into play in the cross-orbital couplings. Such ferromagnetic couplings, 
generated as a secondary effect by the on-site Hund's coupling, were in fact predicted very recently based on an electronic two-orbital model.\cite{raghuvanshi11} 
It would be interesting to also explore the role of effectively ferromagnetic couplings and partially hidden order also for the superconductivity of 
the pnictides.\cite{seo08}

In principle it should be possible to detect the partially hidden aspect of the stripe order in NMR experiments, where different nuclei 
should probe the A and B spins differently and, thus, enable detection of their individual moments. This may require the hyperfine couplings 
to be determined more accurately than presently \cite{kitagawa} (including the different couplings to A and B spins). 

Beyond its relevance in the context of the pnictides, the DSM has a remarkably rich phase diagram, giving access to intriguing multi-orbital magnetic 
effects that are not accessible with the standard single-orbital $S=1/2$ or $S=1$ Heisenberg models. We have here pointed out several phases that had not
been studied previously. It would also clearly be interested to carry out further studies to characterize these states and their excitations. There
may still also exist other states in the large space of parameters of the DSM. 
\null\vskip4mm\null

\acknowledgments

We would like to thank Arnab Sen for stimulating discussions. This research was supported by Grant No.~NSFC-11074310 and the Fundamental 
Research Funds for the Central Universities (DXY) and by the National Science Foundation under Grant No.~DMR-1104708 (AWS). DXY would like to thank 
the Condensed Matter theory Visitors program at Boston University for support and AWS gratefully acknowledges support from the Fundamental Research 
Funds for the Central Universities for a visit to Sun Yat-Sen University.

\end{document}